\documentclass[conference]{IEEEtran}
\IEEEoverridecommandlockouts
\usepackage{amsmath}

\usepackage[dvipsnames]{xcolor}

\usepackage{color}
\usepackage{xspace}
\usepackage{url}
\usepackage{booktabs} % Nicer tables.

\usepackage{array}
\newcommand{\PreserveBackslash}[1]{\let\temp=\\#1\let\\=\temp}
\newcolumntype{C}[1]{>{\PreserveBackslash\centering}p{#1}}
\newcolumntype{R}[1]{>{\PreserveBackslash\raggedleft}p{#1}}
\newcolumntype{L}[1]{>{\PreserveBackslash\raggedright}p{#1}}

\newif\ifsubmit
\submitfalse

\ifsubmit
    \newcommand{\amir}[1]{}
    \newcommand{\hamed}[1]{}
    \newcommand{\saleh}[1]{}
    \newcommand{\helia}[1]{}
    \newcommand{\sina}[1]{}
    \newcommand{\sinaRevHelia}[1]{}
    \newcommand{\amol}[1]{}
    
    \newcommand{\todo}[1]{}
\else
    \newcommand{\amir}[1]{\textcolor{blue}{Amir: #1}}
    \newcommand{\hamed}[1]{\textcolor{cyan}{Hamed: #1}}
    \newcommand{\saleh}[1]{\textcolor{ForestGreen}{Saleh: #1}}
    \newcommand{\helia}[1]{\textcolor{Purple}{Helia: #1}}
    \newcommand{\sina}[1]{\textcolor{Orange}{Sina: #1}}
    \newcommand{\sinaRevHelia}[1]{\textcolor{Apricot}{sinaRevHelia: #1}}
    \newcommand{\amol}[1]{\textcolor{WildStrawberry}{amol: #1}}

    \newcommand{\todo}[1]{\textcolor{red}{TODO: #1}}
\fi

\usepackage{cleveref} % adds \cref{label} command

\crefformat{section}{\S#2#1#3} % see manual of cleveref, section 8.2.1
\crefformat{subsection}{\S#2#1#3}
\crefformat{subsubsection}{\S#2#1#3}

\usepackage{float}
\usepackage{epigraph}
\usepackage{amsmath}
\usepackage{adjustbox}
\usepackage{mathtools, nccmath}
\DeclarePairedDelimiter{\ceil}{\lceil}{\rceil}
\usepackage[utf8]{inputenc}
\usepackage{dirtytalk}
\usepackage{graphicx}
\usepackage{pgfplots}
\usepackage{tikz}
\usepackage{cite}
\usepackage[numbers,sort&compress]{natbib}

\usepackage{graphicx}% Include figure files
\usepackage{dcolumn}% Align table columns on decimal point
\usepackage{bm}% bold math
\usepackage{dirtytalk}
\usepackage[utf8]{inputenc}

\usepackage{float}
\floatstyle{plaintop}
\restylefloat{table}
\usepackage{algorithm2e}
\usepackage{pgfplots}
\usepackage{tikz}
\usepackage{subfigure}
\usepackage{bchart}
\usepackage[font=footnotesize,labelfont=bf]{caption}
\usepackage{graphicx}
\usepackage{subfig}
\usepackage{alphalph}
\usepackage[utf8]{inputenc}
\usepackage{dirtytalk}
\usepackage{subfig}
\usepackage{lmodern}
\usepackage{stackengine}
\usepackage{multirow}

\usepackage{amssymb,amsfonts}
\usepackage{algorithmic}
\usepackage{graphicx}
\usepackage{textcomp}
\usepackage{xcolor}
\def\BibTeX{{\rm B\kern-.05em{\sc i\kern-.025em b}\kern-.08em
    T\kern-.1667em\lower.7ex\hbox{E}\kern-.125emX}}
\begin{document}

\title{Generating High Quality Random Numbers : A High Throughput Parallel Bitsliced Approach\\
 \thanks{Institute for Research in Fundamental Sciences}
}

\author{\IEEEauthorblockN{Saleh Khalaj Monfared}
\IEEEauthorblockA{\textit{School of Computer Science} \\
\textit{IPM$^1$}\\
Tehran, Iran \\
monfared@ipm.ir}
\and
\IEEEauthorblockN{Omid Hajihassani}
\IEEEauthorblockA{\textit{Dep. Electrical and Computer Eng.} \\
\textit{University of Alberta}\\
Edmonton, Canada \\
hajihass@ualberta.ca}
\and
\IEEEauthorblockN{Soroush Meghdadi Zanjani}
\IEEEauthorblockA{\textit{School of Computer Science} \\
\textit{IPM}\\
Tehran, Iran \\
sm@ipm.ir}
\and
\IEEEauthorblockN{\hspace{10mm} Mohammadsina Kiarostami}
\IEEEauthorblockA{\hspace{10mm}\textit{School of Computer Science} \\
\hspace{10mm}\textit{IPM}\\
\hspace{10mm}Tehran, Iran \\
\hspace{10mm}skiarostami@ipm.ir}
\and
\IEEEauthorblockN{\hspace{10mm}Dara Rahmati}
\IEEEauthorblockA{\hspace{10mm}\textit{School of Computer Science} \\
\hspace{10mm}\textit{IPM}\\
\hspace{10mm}Tehran, Iran \\
\hspace{10mm}dara.rahmati@ipm.ir}
 \and
\IEEEauthorblockN{\hspace{15mm}Saeid Gorgin}
 \IEEEauthorblockA{\hspace{15mm}\textit{School of Computer Science} \\
\hspace{15mm}\textit{IPM}\\
\hspace{15mm} Tehran, Iran \\
\hspace{15mm} gorgin@ipm.ir}

}

 %We demonstrate that light-weighted stream ciphers such as MICKEY 2.0 stream cipher algorithm for  

%We depict the capability of light-weighted stream-ciphers such as MICKEY 2.0 in generating high quality random numbers in parallel that satisfy the randomness criteria with [outstanding] performance, while demonstrating our method's scalability and flexibility that allows it to be used with a variety of similar PRNGs.

\maketitle

\begin{abstract}
In this work a high throughput method for generating high quality Pseudo-Random Numbers using the bitslicing technique is proposed. In such technique, instead of the conventional row-major data representation, column-major data representation is employed which allows the bitsliced implementation to take full advantage of all the available datapath of the hardware platform. 
By employing this data representation as building blocks of algorithms, we showcase the capability and scalability of our proposed method in variety of PRNG methods in category of block and stream ciphers. While demonstrating the suitability of stream-ciphers for high throughput PRNG, as an example, we implement and investigate a bitsliced MICKEY 2.0 PRNG by altering the paradigm of internal functions and data structure. The LFSR-based (Linear Feedback Shift Register) nature of the PRNG in our implementation perfectly suits the GPU's many-core structure due to its register oriented architecture and allows the usage of bitslicing technique to further improving the  performance. In our SIMD vectorized fully parallel GPU implementation, each GPU thread is capable of generating a remarkable number of 32 pseudo-random bits in each LFSR clock cycle. We then compare our implementation with some of the most significant PRNGs that display a satisfactory performance in both throughput and randomness criteria.
The proposed implementation successfully passes the NIST test for statistical randomness and bit-wise correlation criteria.
To the best of authors' best knowledge, our method outperforms the current best implementations in the literature for computer-based PRNG and the optical solutions in terms of performance and performance per cost, while maintaining an acceptable measure of randomness. Our highest performance among all of the implemented CPRNGs with the proposed method is achieved by the MICKEY 2.0 algorithm which shows 1.9x improvement over the state of the art NVIDIA's proprietary high-performance PRNG, cuRAND library, achieving 1.6 Tb/s of throughput on the affordable NVIDIA GTX 980 Ti.

\end{abstract}

\vspace{-3mm}
\section{Introduction}

%  [\cite{rahmani2016efficient}, GPU GPGPU \cite{owens2008gpu}]

The emergence of cost-effective, high-performance parallel platforms such as Graphical Processing Units (GPU) and their programmability have allowed researchers across various fields of science and engineering to utilize the specialized processing capability of GPUs to accelerate their computationally demanding applications. GPUs' processing power has been fully leveraged for implementation of machine learning algorithms \cite{rahmani2016efficient, gulli2017deep}, medical image processing \cite{eklund2013medical}, and many other applications.
\par Recently, the high-performance execution on GPU has attracted the attention of many researchers to adapt cryptography problems for execution on massively-parallel GPU platforms \cite{szerwinski2008exploiting, ahmadzadeh2018high}. One problem which is the particular concern of this paper, is the high-throughput generation of sequences of pseudo-random numbers. The high-performance random number generation with an acceptable criterion of randomness is a vital necessity in many computer science disciplines, including stochastic computing \cite{Hojabr2019}, stochastic simulation, i.e. Monte Carlo simulation \cite{binder1993monte}, and cryptography \cite{ahlswede1993common}. 
\par The acceptable criteria for quality of randomness varies across different fields demanding the random number sequence. Perhaps one of the most rigor fields holding very high standards for the randomness is cryptography. In order to showcase the randomness quality our proposed method and namely, our sample implementations are capable of delivering, we employ the criteria used for cryptography purposes. The underlying pseudo-random number generator process, apart from statistical randomness, must accompany other security assurances that vary based on the intended application. The Cryptographically Secure Pseudo-random Number Generator (CSPRNG) processes work based on increasing the entropy of the output sequences which makes the output sequence to be indistinguishable from uniformly random bit sequences. Moreover, the unpredictability of next-bit must be further guaranteed. Here, we intend to apply the bitslicing technique in the software implementation of CSPRNG processes.

%Critical applications like cryptography, in addition to standard statistical randomness properties like uncorrelatedness, [...] have further demanding requirements to decrease predictability of the sequence, including [high entropy]... .
 %increase the difficulty of prediction of the next bits of the random sequence, including [high entropy]... .
%\emph{Cryptographically secure pseudorandom number generators (CSPRNGs)}, a class of \emph{pseudorandom number generators (PRNGs)} that satisfy such requirements are vastly used in 

% [Under more tight conditions for accepting bit streams as random than normal pseudo randomness: cryptographically secure \cite{huang2002hacking}. Need, criteria, ]

In the bitslicing technique, by altering the representation of the input data and computations, we strive to first, increase the utilization of the computation units, and second, reduce the required operations from costly operations to hardware-friendly basic bit-level operations, such as XOR, AND, and OR operations. With the incorporation of the bitslicing technique in our implementations, we can achieve highly-parallel, vectorized execution in the SIMD manner. \cite{biham1997fast} has proposed the successful utilization of the bitslicing technique in the implementation of the Data Encryption Standard (DES) on a 64-bit processor, where the processor is viewed as an SIMD processing unit. In \cite{biham1997fast}, by introducing the bitslicing technique in the implementation of the DES, the 64-bit CPU can be perceived as 64 1-bit CPUs that process 64 chunks of data, simultaneously. In \cite{ahmadzadeh2018high}, the authors have proposed the high-throughput implementation of the bitsliced DES exhaustive key search cryptanalysis technique on programmable GPU platforms.

\par Although, the bitslicing technique has been utilized for cryptanalysis and fast implementation of crypto systems, it has not been used for CPRNGs. A fast, high performance, and reliable CPRNG usually fails to satisfy the criteria in some applications due its complexity compared to regular PRNGs. Here, we would like to improve this drawback for the CPRNGs by incorporating  bitslicing technique and by altering the cryptographich algorithms where performance is improved and the security criteria are maintained. A characteristic of the software implementation of LFSR-based PRNGs is the intrinsic need for repetitive, costly bit-level shift and mask operations. By proposing the bitslicing technique and changing the data and computation representation, we have successfully transformed the aforesaid costly bit-level shift and mask operations to more efficient register swapping techniques. 
%Moreover, with the bitsliced implementation of CSPRNGs, we have reached full utilization of the computational units' processing capability. In the bitslicing representation technique, instead of processing data from one data chunk, each computational unit processes 32 data chunks, simultaneously. This fact, alongside the reduction in the computational complexity of the required operations, greatly improves the execution throughput of our PRNG implementations.
\par As our main contribution, in this paper we provide a comprehensive study on implementation of PRNGs on GPUs by utilizing the bitslicing technique. In order to showcase the the scalability of our proposed method and demonstration of suitability cryptosystem for our purpose, we employ multiple stream and block ciphers, some of which has not been studied before with bitslicing technique. We demonstrate that under the proper choice of the suitable algorithm and by conveniently applying the required necessary step which include steps to apply transformations to functional building blocks and data structures, our proposed method exhibits significant performance in terms of raw computational throughput, throughput per computational power of the device under utilization and also reliability and quality of the generated random numbers by putting it under the NIST test. On top of it all, we present an PSRNG implementation based on Mickey 2. stream cipher that to the best of our knowledge, outperforms all the available implementations of PRNG in terms of performance. In our GPU implementation, our version of Mickey 2 outperforms the Nvidia's proprietary cuRAND, random number generator, by 1.9 X,  despite of its complexity in the algorithm itself.
%as our main contribution, we have proposed a high-performance bitsliced software implementation of the stream cipher Mickey 2.0 pseudo-random number generator. Moreover, we have evaluated the cryptographical security of the generated pseudo-random sequences from our implementation with NIST test, where our implementation has fully passed the requirements and have performed roughly ideally in the remaining constraints. To the best of our knowledge, our implementation outperforms all the available implementations of PRNG in terms of performance. In GPU implementation, our implementation outperforms the Nvidia's proprietary cuRAND, random number generator, by 6.6 X.
To showcase the true capability of the bitslicing technique in software implementation of PRNGs, we have used multiple Nvidia GPUs and performance monitoring and evaluation tools. 
% Although, we have proposed and evaluated our technique on the GPU platform, the bitsliced implementation of pseudo-random number generators can be fruitfully applied to other hardware and software implementations and is not bounded to any kind of platform.
\\The rest of this paper is as follows: In section II, we will introduce a detailed background on the PRNGs, Linear Feedback Shift Registers (LFSR), and some crypto-systems which are employed in our proposed method such as Mickey 2.0. Section III discusses the related efforts to PRNG and RNG implementations. Section IV gives our proposed methodology and elaborates on the incorporation of the bitslicing technique in our implementation along with an examples in different application. Section V, gives the evaluation results achieved from the performance and correctness of our proposed methodology on multiple GPUs. Section VI concludes the paper and discusses future works.

% [Two methods of generation, 
% physical, random and 
% computer based, pseudo random, they fulfill certain criteria \cite{vadhan2012pseudorandomness}]

% for IEEE journal papers produced under \LaTeX\ using
% IEEEtran.cls version 1.8b and later.
% % You must have at least 2 lines in the paragraph with the drop letter
% % (should never be an issue)
% I wish you the best of success.
\section{Background}
\par In this section, we will thoroughly give a background on the random number generation literature, the bitslicing technique and related concepts such as linear-feedback shift register (LFSR) and the underlying mechanisms of employed algorithms for pseudo-random number generation. 

\subsection{Random Number Generation}
Truly random number generator processes are set to be non-deterministic, a condition under which the generated random sequence can not be determined in advance. Truly random number sequences can be generated from sampling of truly random sequences such as physical truly random phenomenon, including \textit{Thermal Noise} \cite{jun1999intel}, \textit{Electrical Noise} \cite{cicek2014novel}, and  \textit{Laser} or \textit{Optical} mechanisms \cite{kanter2010optical}. However, such random number generators that use physical phenomenon are costly. Also, the unavailability of the required apparatus limits the scope of the usage for general applications. Although, such random sequences can be stored for later use which also limits the availability and security in certain applications.
\par The aforementioned issues of cost and availability lead to the use of digital computers in the generation of random numbers. Pseudo-random number generators are not truly random processes which roots from the deterministic essence of digital computers but are specifically designed to meet certain criteria of randomness in their generated sequences. One of the first PRNG methods that uses a random seed and relies on the randomness of the seed for the generation of reproducible random sequences is the Middle Square Method (MSM) \cite{von1963various}. With truly random seeds, PRNGs can generate random sequences until the seed is repeated and the sequence repeats in the output. The size of the initial seed indicates the size of the generated random sequence before the repeat in the generated sequence. \par One feature of PRNG processes is that with the use of the same seed, the generated random sequence can be reproduced which can be exploited in some scenarios such as end-to-end communications. On the other hand, it would also be computationally infeasible to find the random input seed that the PRNG process is using to generate the pseudo-random sequence by exhaustively searching the seed space with a part of the sequence, to find and predict the next-bit of the sequence. To ensure this, the size of the seed must be set to a large enough.
\subsection{Linear Feedback Shift Registers}
\begin{figure}[h]
\centering
\includegraphics[width=0.95\linewidth]{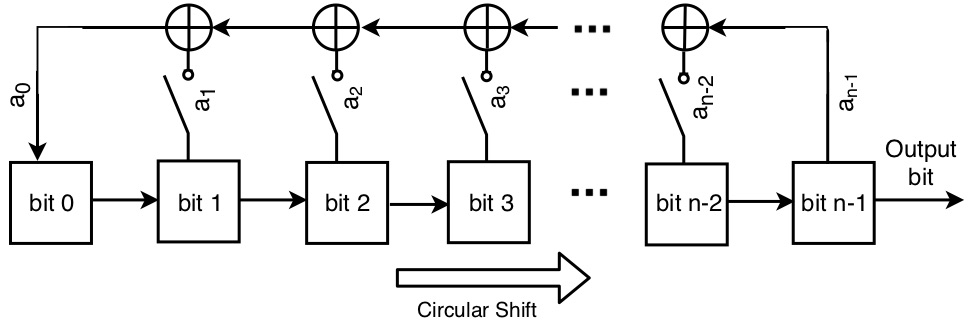}
\caption{A basic n-bit Linear Feedback Shift Register}
\label{lfsr1}
\end{figure}

Based on the mathematical foundation of cyclic codes over finite field of $GF(2)$, the Linear Feedback shift registers have been employed both in software and hardware for a wide range of applications including transmission error checks \cite{koopman200232}, high-performance counters and of course pseudo-random number generators. In LFSR, the feedback tabs which determine the next state of the system if combined linearly could directly impact the input of the system  when the register is shifted at each clock-cycle. Figure \ref{lfsr1} demonstrates a basic representation of a single $n$-bit LFSR. The arrangements of the tabs could be represented by a polynomial referred to as the feedback polynomial.
\vspace{2mm}

For a simple $n$-bit LFSR, the coefficients, the operations defined over $GF(2)$, and the reciprocal characteristic polynomial could be represented as is in Equation \ref{eq:pl}:
\begin{equation} \label{eq:pl} 
\begin{split}
 & p(x)=\sum_{i=0}^{n-1}a_{i}x^{i}; a_{i},x\in GF(2)\\
 & a_{0},a_{n-1}=1
\end{split}
\end{equation}
\par Also, it is worth noting that in many applications, in order to maximize the LFSR period length ($i.e.$ $2^{n}-1$), a primitive polynomial is chosen as the tapping coefficients for the LFSR.

\subsection{Stream Cipher and Block Ciphers}

As already mentioned, cryptographic properties of block and stream ciphers are known to be suitable to generate high quality pseudo random numbers. Among all of various and different proposed stream and block cipher algorithms, here we investigate the two stream and a block cipher which are to be implemented in the bitsliced representation. Some of these ciphers are specifically designed for efficient hardware implementation while guaranteeing acceptable level of security. The ECRYPT Stream Cipher Project (eSTREAM) Profile 2 stream ciphers are particularly suitable for hardware applications with restricted resources such as limited storage, gate count, or power consumption  \cite{babbage2008estream}. We recommend using stream ciphers instead of block ciphers for fast and high performance implementations due to their lightweights architecture. As will be shown in evaluation section, the Mickey 2.0 algorithm shows more promising results compared to block ciphers such as AES.
\subsubsection{MICKEY 2.0 Stream Cipher}
MICKEY 2.0 or Mutual Irregular Clocking KEYstream generator is the second generation stream cipher of the MICKEY family by Babbage and Dodd \cite{babbage2006stream}. Armed with the fact that the MICKEY 2.0 algorithm is inherently light-weight in hardware implementation, its feedback shift register based architecture can be easily implemented with our proposed bitslicing technique.
\par The state machine of the algorithm consists of two 100-bit shift registers, one linear and one non-linear, both clocked irregularly under the control of each other.

\begin{figure}[h]
\centering
\includegraphics[scale=0.45]{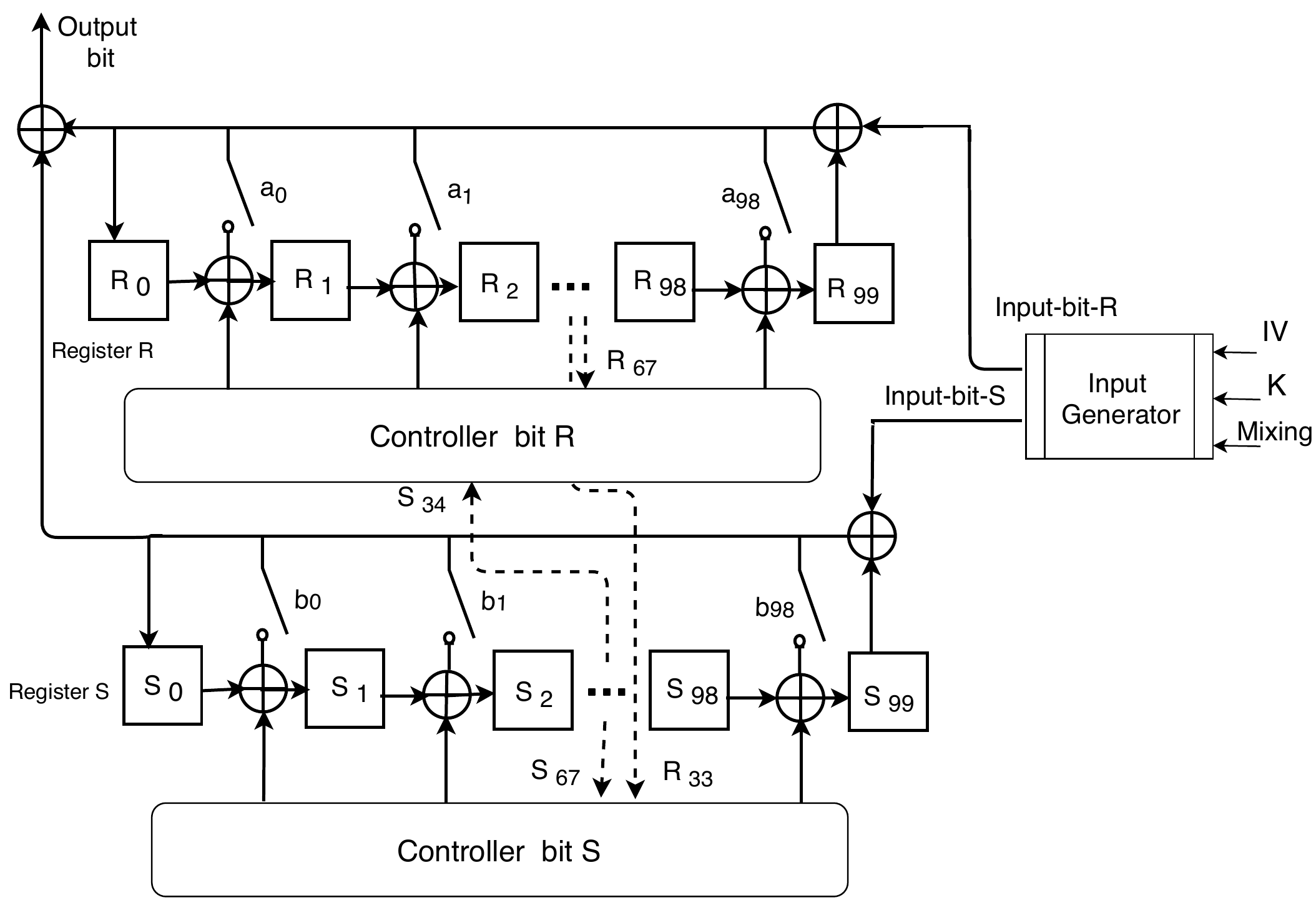}
\caption{Illustration of Mickey 2.0 Stream Cipher Algorithm}
\label{mickey}
\end{figure}

% The MICKEY 2.0 specific clocking mechanisms contribute to the cipher’s cryptographic strength while allowing guarantees on period and pseudorandomness.
It is stated that each key can be used with up to $2^{40}$ different IVs of the same length, and that $2^{40}$ keystreams can be generated from each key/IV pair. Figure \ref{mickey} shows the Galois-based structure of the MICKEY 2.0 algorithm. 
\vspace*{2mm}

\par The MICKEY 2.0 designers have also specified a scaled-up version of the cipher called MICKEY-128 2.0, which employs a 128-bit key and an initialization vector of up to 128 bits. It is stated in the specification that the irregular clocking mechanism makes the parallel implementation somehow \textit{not so straightforward}. However, as will be thoroughly investigated later, our proposed bitsliced algorithm utilizes a fully parallel implementation of MIKCEY.
Furthermore, it has been noted by Gierlichs et al. \cite{gierlichs2008susceptibility}, that straightforward implementations of the MICKEY ciphers are likely to be susceptible to timing or power analysis attacks. However, the system could be immunized by software techniques like masking, making these attacks significantly ineffective. Otherwise, there have been no known cryptanalytic advances against MICKEY 2.0 or MICKEY-128 2.0 after its publication in eSTREAM.\cite{rukhin2001statistical}

\subsubsection{Advance Encryption Standard}
Advanced Encryption Standard also known as AES is the the most famous and used block cipher in communication today. After five years of competition and standardization, National Institute of Standards and Technology (NIST) selected Rijndael block cipher to supersedes Data Encryption Standard (DES) in 2001 as AES \cite{rijmen2001advanced}. NIST AES specification of AES introduces three versions of  Rijndael cipher with 10, 12, and 14 rounds of ciphering with 128, 192, and 256 bits of keys, respectively.
The AES algorithm is consist of three major building blocks and is processed in byte granularity in extended Galois Field of $GF(2^8)$. The state matrix of 16 bytes in a $4\times4$ matrix is constructed each step is iterated in AES. S-box or the substitute byte is the only non-linear part of the AES which is a simple substitutions table and responsible for a non-linearity in the cipher. The S-box is usually implemented by look-up table in memory in software and hardware implementations. However, for our proposed method, the S-box is efficiently implemented by bit-level gates. The Mix-Column and Shift-rows boxes are responsible for diffusion of the data in cipher. The Mix-Column is a Galois-based Matrix Multiplication step and the Shift-Row is a simple linear byte swapping in the rows of the state matrix. Our proposed method for PRNG in the AES is based on the CTR (Counter Mode) algorithms discussed in \cite{FastAES2019}. 
% \subsection{Statistical Tests for Randomness}
% The statistical randomness tests a
% Come on Omid boi
\subsubsection{Grain Stream Cipher}
\begin{figure}[h!]
\centering
\includegraphics[scale=0.5]{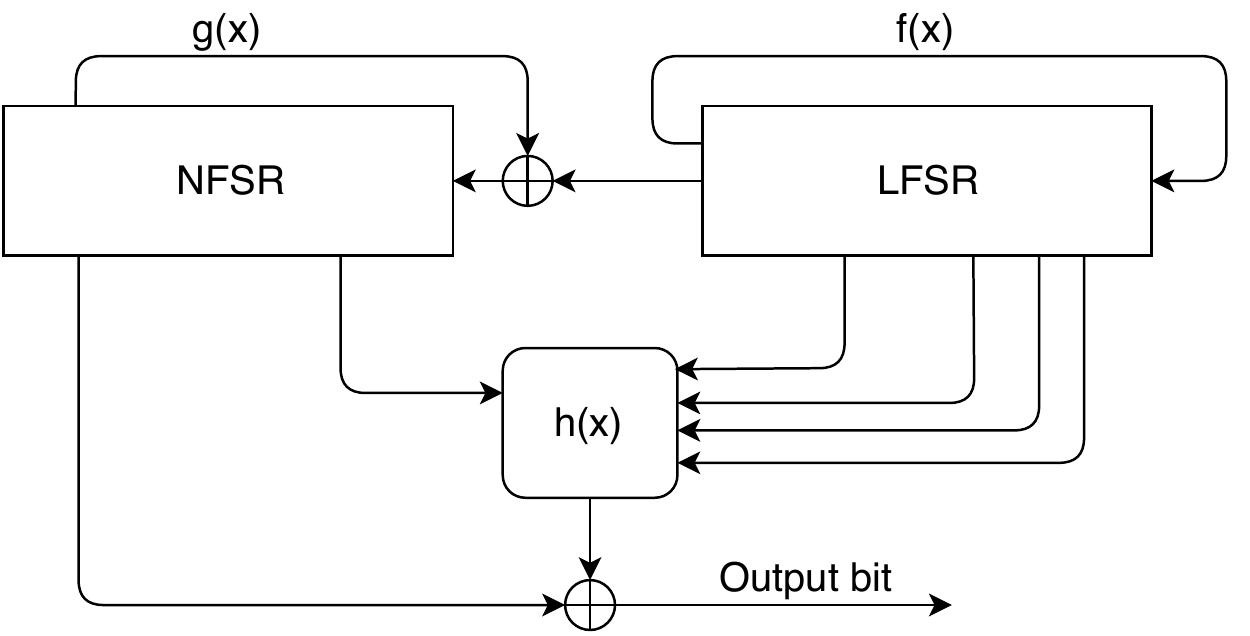}
\caption{Illustration of Grain Stream Cipher Algorithm}
\label{grain}
\end{figure}

Grain is also a winner of eSTREAM portfolio for Profile 2, specifically designed for restricted hardware environments. The Grain developed by Hell et al. \cite{hell2007grain}, is constructed by two 80 bits Linear and Non-Linear Feedback Shift Registers (NFSR) which are shifted together at each clock-cycle. The NFSR is controlled by a feed-back function of itself and LFSR output. Similarly, the LFSR is also controlled by feedback function. The cipher is normally initialized with 80-bit key and a 64-bit Initial Vector(IV) which is directly fed into the NFSR and LFSR respectively at the beginning. The specification recommends a 160 clocks of initialization before the key stream generation. The light-weighted architecture of the Grain structure couple with the shift-registers used in this algorithm, makes it a great nominees for the bit-sliced implementation. A high level demonstration of Grain stream Cipher is given in Figure \ref{grain}.

% \subsubsection{Extended Tiny Encryption}

\section{Related Efforts}
% Come on Omid boi
%Sina Boi coming:
%Since Von to now (History)

% \epigraph{Anyone who considers arithmetical methods of producing random digits is, of course, in a state of sin.}{\textit{John von Neumann, 1951
% }}
% In a wide range of applications, pseudo random numbers are been generated in physical or arithmetical ways. In 1951, Von Neumann said "Anyone who considers arithmetical methods of producing random digits is, of course, in a state of sin. " and claimed that physical methods maybe have eventuated wrong which need to be checked \cite{von195113}.  

% My Archive:

%Pseudo Random Number Generators (PRNGs) are widely used in a variety of applications such as encryption, simulation, gaming, art and communications to name just a few examples. A good quality random number generator is an asset for applications like encryption, randomized designs and network and information security[True random number generator using GPUs and Histogram equalization techniques1] 2011

% in these cryprographic applications: Exhaustive Search, Time-Memory Trade-Off (TMTO) and Counter Mode of Operation using LFSR has two main advantages: Time and Area. [Application of LFSRs for Parallel Sequence Generation in Cryptologic Algorithms]  2006

% Cryptographic systems should use only true random number generators for producing keys and other secret quantities [A Design of Reliable True Random Number Generator for Cryptographic Applications] 1999

%

%fpga asic,... then gpu and specially optic shits
%finally cuRand

% [Soroush]
Random number generation has been a topic of interest for researchers and developers for decades. Numerous theoretical and practical studies have investigated the complexity of generating high quality random numbers and evaluating them \cite{l1990random,mascagni2000algorithm}.
Making use of parallel platforms for acceleration of RNGs has also been a matter of consideration in the literature \cite{mascagni1999sprng}. Staring around the 2000's, the emergence of general-purpose computing on graphics processing units has opened new horizons for high-performance generation of random numbers. In 2006, M Sussman et al. published one of the first works utilizing the power of GPGPU on the subject of RNG \cite{sussman2006pseudorandom}. In the years to follow, many researchers and developers reported successful implementations of PRNGs with increase in performance on parallel platforms, achieving remarkable speedups over the CPU platforms  of their time and outperforming similar efforts \cite{gong2010accelerating}. In the subject of high-performance parallel PRNG, the performance of Nvidia’s proprietary PRNG cuRAND library \cite{cuRAND} has always been a forceful competition, still in some cases researchers have reported their work excelling the performance of the cuRAND of their time \cite{nandapalan2011high}. Compared to the most significant recent efforts on pseudo-random number generation on GPU \cite{teh2015gpus, al2018acoustic}, the cuRAND library seems to remain the major dominating player in the field with performances brighter than competitors in terms of both throughput and throughput per processing power of the device under utilization.
\par Regardless of all the performance advancements in the HPC community, there have always been skepticism and critical opinions regarding arithmetic methods for generating random numbers. Highly favored from physics academic community \cite{li2018ultrafast} quote by Von Neumann stating \say{Anyone who considers arithmetical methods of producing random digits is, of course, in a state of sin} \cite{von1963various} is famously cited to designate the difference between natures of generated random numbers by physical and computational methods. Although, not truly random numbers by definition, computationally generated random numbers have gained serious attention in recent years due to their accessibility, affordability, and their ability of employment in high-performance platforms thanks to the development in parallel hardware devices. Theoretical and practical methods regarding generation and evaluation of computationally generated random numbers have advanced so far that allows these arithmetically generated random numbers to be securely utilized in sensitive applications like cryptography.
\par It is worth noting that the quality of random numbers is dependent on its target user and application domain \cite{l1990random}. Trivial computer games and sensitive cryptographic applications have different requirements and criteria for measurement of how \say{good} a random number is. Therefore to ensure that the randomness properties of a sequence is satisfactory for a certain application, the measurement of the quality of the generated sequence is of high significance. Consequently, efforts have been made to develop procedures and tools capable of evaluating the desirable statistical properties of a given random sequence \cite{l2007testu01}. NIST SP 800-22 \cite{rukhin2001statistical} is a statistical test suite from the national institute of standards and technology (NIST) designed to probe RNGs for both statistical and cryptographical properties, ensuring the qualification of the passing RNG for its target usage in cryptographic applications.
\par Taking advantage of the GPU platform in generation of random numbers in comparison to the physical and the optical methods
% [https://arxiv.org/pdf/1712.02254.pdf] 
\cite{kanter2010optical,liu2018device,xu2019spad}, other hardware platforms \cite{thomas2009comparison} such as FPGA \cite{stanchieri2019true}
and CPU \cite{matsumoto1998mersenne} enables the users to strike a balance between the quality of randomness, flexibility, obtainability, affordability, performance, and outstanding performance per cost metrics.
\begin{table}[h]
\begin{adjustbox}{width=\columnwidth,center}
\centering
\caption{Priviously proposed PRNG Implementaions on GPU}
\label{tab:GPUspecs}

\begin{tabular}{|c|c|c|l|c|c|l|}
\hline
Ref                         & \textit{Year} & \textit{GPU} & \textit{GPU's GFLOPS} & \textit{Method} & \textit{Method's Gbps} & \textit{Normalized (Gbps/GFLOPS)} \\ \hline
\cite{langdon2008fast}      & 2008          & 8800 GTX     & 345.6               & RapidMind       & 26                   & 0.0752314814814815                \\ \hline
\cite{pang2008generating}   & 2008          & 7800 GTX     & 20.6                & CA-PRNG         & 0.41                 & 0.0199029126213592                \\ \hline
\cite{langdon2009fast}      & 2009          & T10P         & 622.1               & ParkMiller      & 35                   & 0.0562610512779296                \\ \hline
\cite{gong2010accelerating} & 2010          & S1070        & 2488.3              & N/A             & 4.98                 & 0.00200136639472732               \\ \hline
\cite{nandapalan2011high}   & 2011          & GTX 480      & 1344.96             & xorgensGP       & 527.5                & 0.392204972638591                 \\ \hline
\cite{gao2013gasprng}       & 2013          & GTX 480       & 1344.96             & GASPRNG         & 37.4                 & 0.0278075184392101                \\ \hline
\end{tabular}

% \begin{tabular}{|c|c|c|c|c|}
% \hline
% Ref                                          & \textit{Year} & \textit{GPU}       & \textit{Method} & \textit{Max Throughput (Gbps)} \\ \hline
% \cite{langdon2008fast}      & 2008          & 8800 GTX           & RapidMind       & 26                             \\ \hline
% \cite{pang2008generating}   & 2008          & 7800 GTX           & CA-PRNG         & 0.41                           \\ \hline
% \cite{langdon2009fast}      & 2009          & T10P               & ParkMiller      & 35                             \\ \hline
% \cite{gong2010accelerating} & 2010          & C1060/S1070        & N/A             & 4.98                           \\ \hline
% \cite{nandapalan2011high}   & 2011          & GTX 480/GTX 295    & xorgensGP       & 527.5                          \\ \hline
% \cite{gao2013gasprng}       & 2013          & GTX480/C1060/M2070 & GASPRNG         & 37.4                           \\ \hline
% \end{tabular}

\end{adjustbox}
\end{table}
\par Unfortunately, current methods of random number generation on GPU do not take full advantage of this hardware platform. Latest efforts on CSPRNG on high-end GPUs perform poorly in utilizing the massive parallelization capabilities of GPU in order to reach high generation performances \cite{al2018acoustic}. Table I represents claimed performance of some of the related efforts along with the processing power of the GPU on which they could reach their peak performance. Also, to establish a fair platform for comparison, peak performance achieved by each method was normalized to the processing power of their employed device and reported as well.
The performance achieved by the state of the art RNGs such as Nvidia’s proprietary cuRAND library still does not fully exhibit the full potential of modern GPUs. We show this performance can be further enhanced while maintaining a reasonable cryptographically properties via applying the bitslicing approach to the implementation of GPU-based CSPRNGs.

\section{Proposed Method}
In this section, we propose the implementation of the bitslicing software technique for high-throughput cryptographically safe pseudo-random generation which is based on cryptographic algorithms. The approach of column-major bitsliced data representation scheme is firstly introduced and thoroughly discussed and their advantages compared to the na\"ive implementation are explained. Afterward, by incorporating the bitslicing technique into the implementation of our work, a high-throughput algorithm for LFSR architecture is presented and then, as another application a cyclic redundancy check (CRC) using the proposed architecture is described.  The implementation of parallel MICKEY algorithm as an example for bitsliced stream ciphers is presented for Random Bit Generation (RBG). Finally, further GPU optimization techniques used in our implementations are discussed. It it also worthy to mention that we have implemented the bitsliced version of three cryptosystems, namely \textit{AES}, \textit{MICKEY 2}, and \textit{Grain} algorithms as CPRNGs to show the extensiveness of the proposed method. However, here only the alteration MICKEY 2 algorithm is described as an example.
\subsection{Bitsliced SIMD Vectorization and Data Representation}
Bitslicing technique was employed by Biham \cite{biham1997fast} for implementation of cryptographic algorithms. At the time, the technique was able to speedup the previous implementations of the Data Encryption Standard (DES) to accelerate the exhaustive search procedure. As mentioned, by the emergence of high-performance, affordable general purpose GPUs, the bitslicing technique has been successfully employed by many works such as \cite{nishikava2017}, \cite{FastAES2019}, and \cite{ahmadzadeh2018high} as a software solution for high-throughput demanding cryptographic applications on GPUs.
% Recently, authors in \cite{FastAES2019} have proposed a bitsliced implementation of the Advanced Encryption Standard (AES) algorithm based on GPU architecture.
This bitslicing-based implementation, leveraged by the column-major data representation, reaches an unprecedented throughput of Terabits per second (Tbps) both in encryption and decryption.
\par Here, before getting into the details of the bitsliced implementation of the proposed PRNG via MICKEY 2.0, we discuss the data representation scheme employed in our work.
% which could lead to a remarkable performance for random bit generation.
The proposed representation scheme fits the parallel architecture of the GPU and fully utilizes the available datapath of the computational units in the deployment hardware. 
% The proposed representation scheme not only perfectly suits the parallel architecture of the GPU, but also, fully utilizes the available datapath of the computational units in the deployment hardware. 
\par Our proposed data representation scheme, uses column-major data representation, instead of the conventional row-major representation. By the row-major representation, we refer to the representation used in the way that the data is stored in common programming practices. In our implementation, we store state bits and other supplementary and temporary registers in the column-major representation. By doing so, we strive to achieve full SIMD execution of a number of 32 (in the case of single precision calculations) bits from different data chunks at each execution clock cycle. In the case of our LFSR implementation, a batch of 32 bits data stored in a single register, represents state bits from 32 uncorrelated different parallel LFSRs having the same bit significance. Hence,

% As shown in Figure \ref{datarepresentation}, 

the first step is to alter the representation to the column-major data representation.
\par For a simple LFSR implementation operating in the conventional row-major representation, one or more registers are used to store the state bits of the LFSR algorithm. Hence, in order to store the n-bit LFSR states with a primitive polynomial (feedback polynomial) of     $p(x)=\sum_{i=0}^{n-1}a_{i}x^{i}$, a number of $\ceil{\frac{n}{m}}$ registers are needed to store LFSR state bits. For instance, for a simple 20-bit LFSR, assuming single precision operations, a single register of 32-bit width is employed to handle the computation of the LFSR state machine.
\par As thoroughly investigated in the previous section, the shift operation is inherent to the LFSR architecture, and in the conventional na\"{\i}ve implementation, costly bit-level shift and mask operations are mandatory at every single rotation of the LFSR state machine. This would considerably limit the overall performance of the RNG circuit, since these bit-access operations should be executed at each rotation. Moreover, in some scenarios, the register utilization in terms of datapath width of the platform cannot be maximized due to the unused number of bits in the conventional row-major data representation. However, our proposed column-major bitsliced data representation, not only compensates for the aforementioned shortcomings inherent to the common practice na\"{\i}ve implementation, but also maximizes the utilization of the processing units in the GPU.

\subsection{Bitslicing Approach Applications : CRC Example }

\begin{figure}[h]
\centering
\includegraphics[width=1\linewidth]{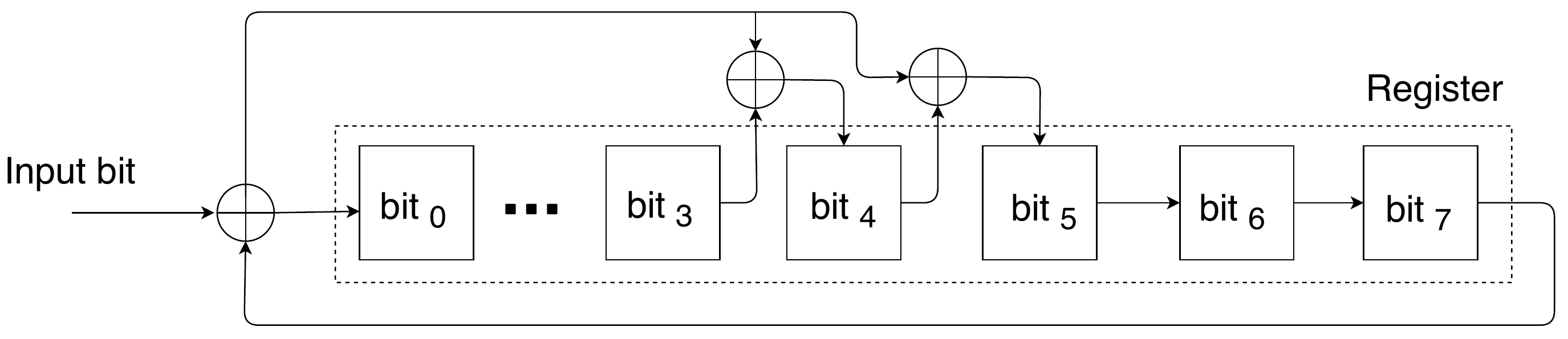}
\caption{A simple CRC-8 example. Na\"{\i}ve implementation}
\label{CRC1}

\end{figure}

As indicated, bitslicing technique could be employed in register-based processors in many applications. In this context, as an example, we examine the usage of column data representation in a simple 8-bit Cyclic Redundancy Check (CRC-8) to show the extensibility of the discussed method. CRC are used to check the error in communication channels with wide range of applications in Wireless Mobile Networks, Wired Ethernet, and countless other applications. As shown in Figure \ref{CRC1}, a simple 8-bit CRC is constructed  by shift-registers and the state value of the CRC is changed by input stream at each cycle. Typically, a CRC is implemented on a single register and the computation is handled by simple shift and mask operation within the register. The CRC output of a specific input data is the final state bits stored in the register which is used to check the correctness of the original data. 

\begin{figure}[h]
\centering
\includegraphics[width=0.95\linewidth]{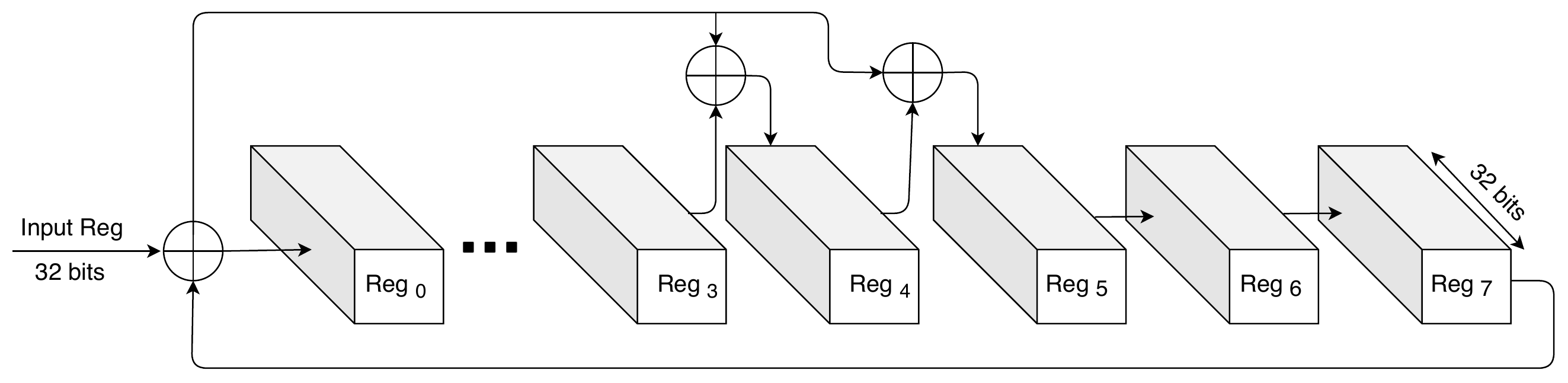}
\caption{A simple CRC-8 example. Na\"{\i}ve implementaion}
\label{CRC2}
\end{figure}

By Taking advantage the bitslice data order one can implement the CRC-8 as shown in Figure \ref{CRC2}. Considering a processor with 32-bit register, this representation constructs a fully paralleled CRC calculation for 32 different data streams, simultaneously without any computational overhead. The shift and mask operation are completely removed and replaced with trivial register reference swapping.

\subsection{Bitsliced LFSR Implementation}
\begin{figure}[h]
\centering
\includegraphics[width=0.95\linewidth]{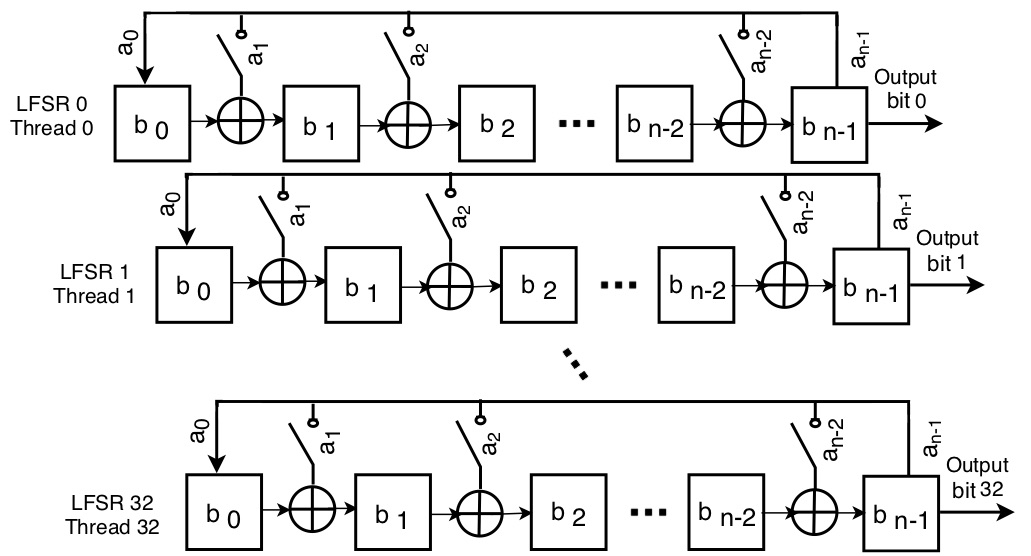}
\caption{Number of 32 parallel LFSR modules executed in 32 threads}
\label{parallellfsr}
\end{figure}

Along with the fact that the costly shift and rotation operations can be further reduced to simple register swapping operations in the bitsliced data representation, here, we investigate the underlying architecture of our proposed bitsliced LFSR implementation on GPU which will be employed for the CPRNGs such as MICKEY 2.0 algorithm.
Moreover, we indicate the properties and advantages of our proposed bitslicing technique accompanied by the column-major data representation. As shown in Figure \ref{CRC2} and explained earlier the LFSR conventional implementations suffer from heavy bit-level shift and mask operations. For fair comparison and for the sake of simplicity, consider the na\"{\i}ve implementation of 32 parallel LFSRs governed by the primitive polynomial $g(x)$ shown in Equation \ref{eq:gx}. Note that there are at least $k$ number of feedback paths in the LFSR algorithm.
    % \bigl| \dot{r} \bigr| & =
    % \sqrt{\medmath{ -\!\bigl(e^{-t} ( \cos t + \sin t )\bigr)^2 + \bigl(e^{-t} (\cos t + \sin t)\bigr)^2 + \bigl(-e^{-t}\bigr)^2}} \\
    %      & = \sqrt{ -\!\bigl(e^{-t} ( \cos t + \sin t )\bigr)^2 + \bigl(e^{-t} (\cos t + \sin t)\bigr)^2 + \bigl(-e^{-t}\bigr)^2}

\begin{align}
  & g(x)=\bigoplus_{i=0}^{n-1}a_{i}x^{i}; a_{i},x\in GF(2) \nonumber \label{eq:gx} \\
   &|A|=k \\
   &A=\{a_{i} | a_{i}\neq{0}\}\nonumber
\end{align}
%   here are a bunch of tests for randomness that are used in the literature in addition to nist
%   figured might be iseful in evaluation:
%   https://en.wikipedia.org/wiki/TestU01
%   https://en.wikipedia.org/wiki/Kolmogorov%E2%80%93Smirnov_test
%   https://en.wikipedia.org/wiki/Diehard_tests
%   see here for instance: 
%   http://cdn.iiit.ac.in/cdn/cstar.iiit.ac.in/~kkishore/lspp12.pdf
As is illustrated in Figure \ref{parallellfsr}, each of these parallel LFSRs are handled by a single thread which results in the execution of 32 parallel threads. Hence, to generate a total number of $M$ pseudo random bits, each LFSR module should be shifted for $M/32$ times where a number of $32\times k$ bit-level XOR operations are needed. 
\\ It is worth noting to know that to use parallel LFSRs in this manner, the shift-registeres should be carefully initialized to eliminate any statistical correlation between the LFSR state machines when the output is not mixed (it is highly recommended to use non-linear mixing before generating the bit stream). Moreover, from the cryptanalysis point of view, the secure threshold for the repeat period (not $2^{n}-1$ in this case) of the employed parallel system should be estimated.
\vspace*{5mm}
\par
\begin{figure}[h]
\centering
\includegraphics[width=0.95\linewidth]{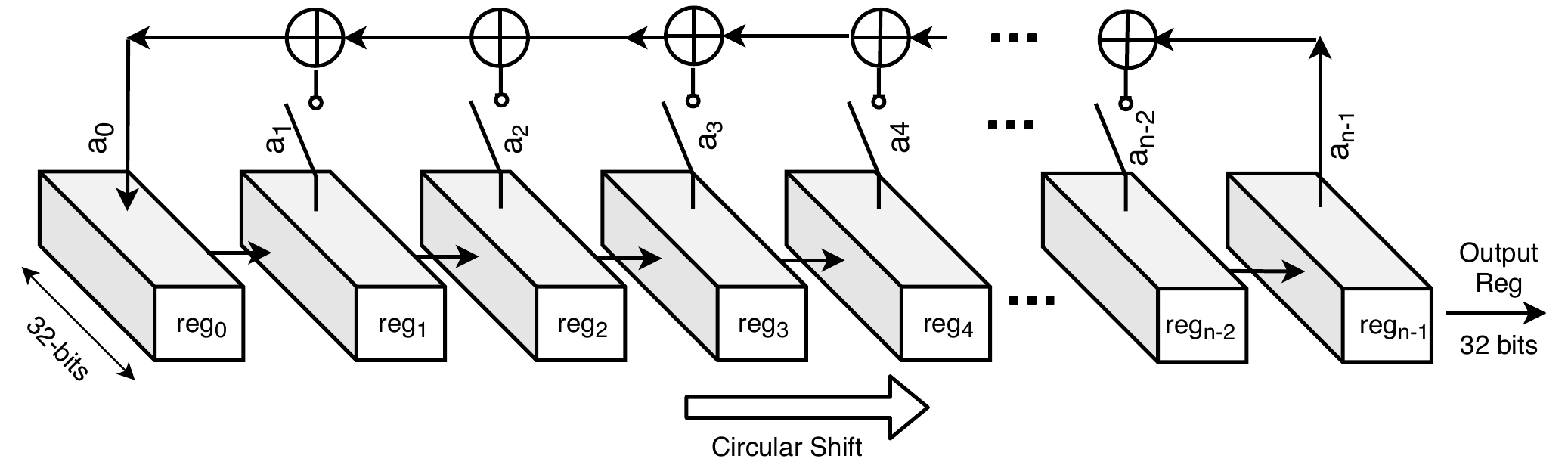}
\caption{The execution of a 32-bit bitsliced LFSR by a single thread}
\label{btclfsr} 
\end{figure}

 Considering the same scenario for pseudo-random bit generation by the use of LFSR, in Figure \ref{btclfsr} the bitsliced LFSR implementation with our proposed column-major representation is demonstrated. Compared to the previous conventional model, to generate $M$ bits in this proposed methodology, the same number of $M/32$ LFSR shift cycles are required. However, this procedure could be executed by a single thread. Also, it is worth mentioning that the $32\times k$ number of costly bit-wise XOR operations (needed at each cycle) is reduced to $k$ number of full-width XOR operations. These operations maximize the datapath utilization. Moreover, as shown in Figure \ref{btclfsr}, the costly bit-level shift operations are replaced by cheaper and more trivial register swapping operations which can be easily done by changing the references of the registers in the software code. Although, changing references in code might be a burdensome task, it greatly reduces the number of needed instructions in the code. Similarly, in this case the registers should be safely initialized from the perspective of cryptanalysis and the period of the usage should be considered. Note that to maximize the repeat period of the LFSR algorithm for PRNG, it is recommended to choose an LFSR with a higher $n$ value.
\vspace{2mm}
\subsection{MICKEY 2.0 Bitsliced Implementation}
\begin{figure}[h]
\centering
\includegraphics[width=\linewidth]{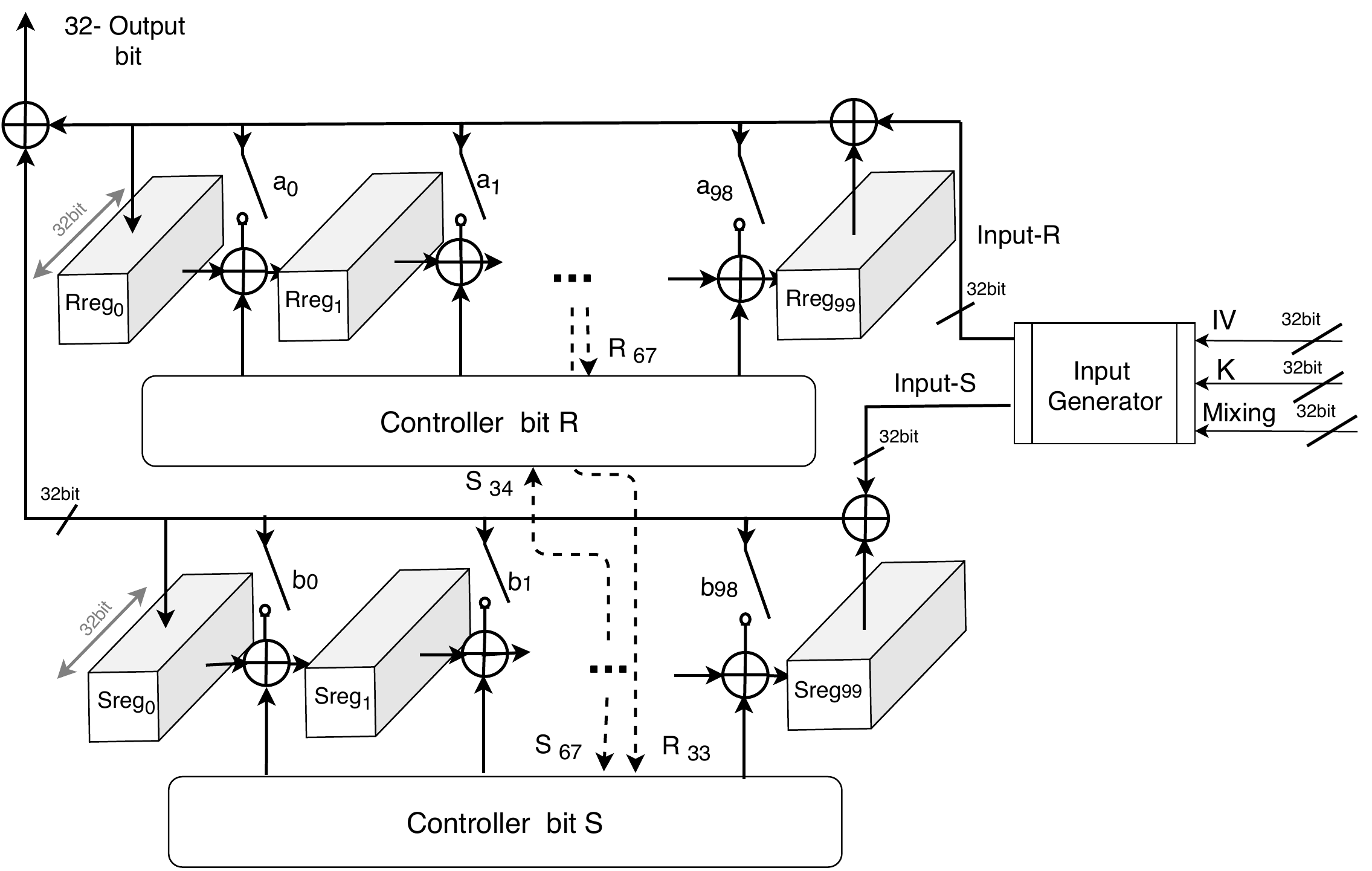}
\caption{32-bit Bitsliced implementation of Mickey 2.0 stream cipher algorithm}
\label{btcmickey}
\end{figure}

For the sake of explanation, here we describe the Mick
As explained, MICKEY 2.0 stream cipher is comprised of two 100-bit registers, namely $S$ and $R$ registers. By incorporating the bitslicing technique into the implementation of the MICKEY 2.0 algorithm, instead of two 100-bit registers, the data representation is altered into column-major order and 200 registers each containing 32 bits are employed. Note that our implementation utilizes single precision computation which occupies 32-bit registers. By doing so, 32 parallel Mickey stream ciphers are executed simultaneously. Figure \ref{btcmickey} demonstrates the parallel bitsliced Mickey architecture. $Rreg_{i}$ and $Sreg_{i}$ represent the $i^{th}$ bits of the $R$ and $S$ registeres in the bitslicing manner, respectively. Each of these registers stores 32 different bits of the same significance for 32 parallel LFSRs modules. Hence, in our implementation each GPU thread is capable of executing 32 parallel Mickey 2.0 ciphers and 32 random bits are generated by each thread at each clock cycle. Also, note that here, the XOR operation is executed on two 32-bits registers and the register-based operations are fully utilized compared to the na\"{\i}ve implementation.
\par To securely and properly initialize our bitsliced Mickey algorithm, we employ a non-linear function to expand a carefully selected pre-stored random number set which generates an 80-bit Initialization Vector (IV) for each MICKEY module ($32\times80$ bits of IV for each thread). It is worth noting that the controller bit functions are designed in the bitsliced representation to calculate all the 32 bits of the controller bits for the feedback procedure. This bitsliced controller is fully optimized to compute the underlying parameters responsible for feed-back procedure in the algorithm. Moreover, the input handler is also executed in 32-bit width mode with no additional overhead and is fully optimized in terms of datapath width.
% \subsection{Shared Memory and Coalesced Access}

\vspace{10 pt}

\section{Evaluation}
In this section, we will present the evaluation results of the performance and the performance per cost metrics achieved from the execution of our proposed CPRNG implementations on a number of different CUDA-enabled GPUs. In this study, six Nvidia GPUs are deliberately selected for evaluation purposes. The employed GPUs each have different structural characteristics such as different single and double precision throughputs and memory bandwidths. These features are carefully selected to represent a wide range of execution platforms. We selected these GPU platforms because of the fact that firstly, the range of the selected GPUs completely represents the platforms available to a wide range of users spanning home and enterprise users. Secondly, these GPUs give us a fair comparison with the previously proposed methods. Moreover, we demonstrate the randomness robustness and reliability of the generated bits by discussing the NIST statistical test results for our sample implementation.
\subsection{Setup}
This section elaborates on the specification of the hardware platforms used for the evaluation of our proposed method. GPU platforms GTX 480, GTX 980 Ti, and GTX 1050 Ti on systems with two Intel XEON E5 2697 V3 CPUs clocked at 2.6 GHz and 128 GB of DDR3 RAM were used for the evaluation process. Moreover, to prove the scalability of the proposed method Tesla V 100 and GTX 2080 Ti GPUs are also utilized for evaluation which are employed in a Virtual Machine environment with the same virtually dedicated specifications. Table II shows the specification details of the aforesaid GPU platforms in terms of the processing power and the memory bandwidth.
\begin{table}[h]
\resizebox{\columnwidth}{!}{%
\begin{tabular}{|c|c|c|c|}
\hline
GPU     & \begin{tabular}[c]{@{}c@{}}Single Precision\\ (GFlops)\end{tabular} & \begin{tabular}[c]{@{}c@{}}Double Precision\\ (GFlops)\end{tabular} & \begin{tabular}[c]{@{}c@{}}Memory Bandwidth\\ (GB/s)\end{tabular} \\ \hline
GTX 480     & 1344                                                                & 168                                                                 & 177                                                               \\ \hline
GTX 980 Ti  & 5632                                                                & 176                                                                 & 337                                                               \\ \hline
GTX 1050 Ti  & 1981                                                                & 62                                                                 & 112                                                               \\ \hline
GTX 1080 Ti & 10609                                                                & 332                                                                  & 484                                                               \\ \hline
Tesla V100 & 14028                                                               & 7014                                                                 & 900                                                               \\ \hline
GTX 2080 Ti & 11750                                                               & 367                                                                 & 616                                                               \\ \hline
% 2080 Ti & 11750                                                               & 367                                                                 & 616                                                               \\ \hline
% V100    & 14028                                                               & 7014                                                                & 900                                                               \\ \hline
% P100    & 9340                                                                & 4670                                                                & 732                                                               \\ \hline
\end{tabular}
\caption{Specification of the GPU platforms used for evaluation}
}
\label{GPUspecs}
\end{table}
\vspace{-3mm}
\subsection{Performance}

\begin{figure}[t]
\centering
\includegraphics[width=1\linewidth]{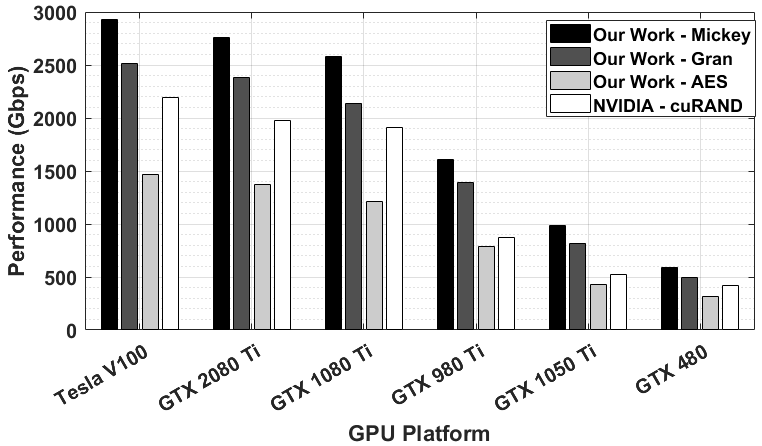}
\caption{Comparing the performance of the proposed method on different GPU platforms}
\label{perf1}
\end{figure}

Figure \ref{perf1} illustrates the achieved performance for our proposed method based on three cryptosystems (AES, Mickey, Grain). In this Figure, we have compared our results with NVIDIA's cuRAND library (on the same platform) since all other previously proposed methods have failed reach the cuRAND performance in PRNGs. The best result obtained on GPU V100 is acquired in the following manner of executing the implemented CUDA kernel code with fixed parameters of \textit{thread blocks} and \textit{thread per block} set to 64 and 256, respectively.
% \vspace{2mm}
The loop size of the code is varying between 4,400 to 13,000, yielding to a different performance throughput. The cuRand results here  are evaluated using the Mersenne Twister algorithm as the default \textit{cuRand} method for RNG. Note that the peak AES performance is more limited compared to the stream ciphers here. This is mainly caused by the complex bitsliced S-box in the AES. Also, the LFSR based structure of the stream ciphers are more compatible with the proposed bitslice technique.
\begin{figure}[H]
\centering
\includegraphics[width=\linewidth]{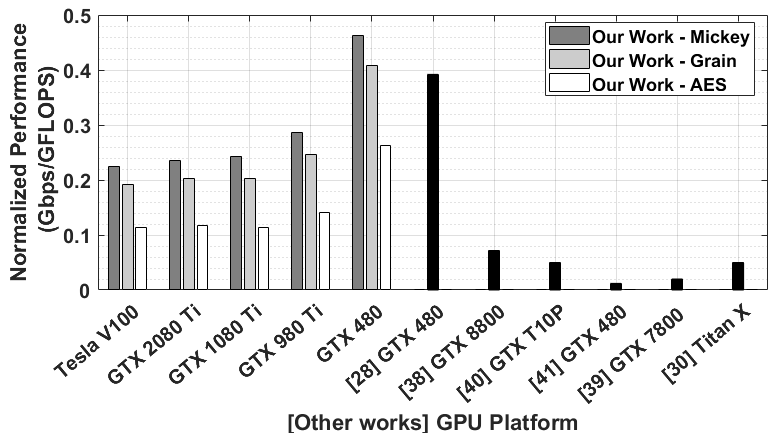}
\caption{Comparison of  normalized performance of the proposed method with previous works}

\label{perf2}
\end{figure}
\subsection{Normalized Performance Evaluation}
\vspace*{-1.2mm}
Due to the lack of access to some of the GPU platforms used in previous works that are currently outdated platforms and to deliver a fair comparison, we follow the method of normalizing the results of our proposed method and related works on parameters of performance per processing power which is shown in Figure \ref{perf2}. However, as already indicated the most important  available library to compare, is the cuRAND which has been considered in our evaluation.
\subsection{Statistical Tests}
We use the NIST SP 800-22 version sts-2.1.2 for testing the statistical and cryptographical properties of our generated random sequences in the Mickey algorithm. Bitstreams generated by our implementation successfully pass all statistical tests. As recommended by the NIST tests, the items are executed using 1,000 instances of 1Mb of random bits generated by our solution. Note that the \textit{significant value} here is considered to be $\alpha=0.01$, and the results of $P-Value$ verify the randomness of the input bitstream.
\begin{table}[!hbt]
 \resizebox{\columnwidth}{!}{%

\begin{tabular}{|c|c|c|c|}
\hline
Test                    & P-value   & Proportion & Result  \\ \hline
Frequency               & 0.251741  & 0.9982      & Success \\ \hline
BlockFrequency          & 0.350485  & 0.9947      & Success \\ \hline
CumulativeSums          & 0.4766135 & 0.9751     & Success \\ \hline
Runs                    & 0.534146  & 0.9781      & Success \\ \hline
LongestRun              & 0.350485  & 0.9562      & Success \\ \hline
Rank                    & 0.213309  & 0.9950      & Success \\ \hline
FFT                     & 0.534146  & 0.9971      & Success \\ \hline
NonOverlappingTemplate  & 0.4821360 & 0.9885  & Success \\ \hline
OverlappingTemplate     & 0.739918  & 0.9912  & Success \\ \hline
% Universal               & X         & 0.8570  & Success \\ \hline
ApproximateEntropy      & 0.350485  & 0.9721 & Success \\ \hline
% RandomExcursions        & X         & 0.9926  & Success \\ \hline
% RandomExcursionsVariant & X         & 0.9966  & Success \\ \hline
Serial                  & 0.7227795 & 0.99982      & Success \\ \hline
LinearComplexity        & 0.739918  & 0.9840    & Success \\ \hline
\end{tabular}
\caption{Evaluation results of NIST statistical suite. The results are the average of 1,000 samples of 1,000,000 bit streams of random numbers generated by the proposed method.}
}
\label{tab:GPUspecs}
\end{table}

\section{Discussion \& Conclusion}
% []
In this work, we propose a high-throughput full parallel cryptographically secure pseudo-random number generator using the bitslicing technique. In this technique, the data from the conventional row-major representation is altered into column-major representation for the purpose of full utilization of the computation datapath of the employed device. Using the bitslicing technique on the LFSR-based cryptographically secure MICKEY 2.0 stream cipher along with other crypto-systems are implemented. This allows of high-performance random number generation by the elimination of the shift and mask operations. Various supplementary techniques such as utilization of shared memory and coalesced memory accesses are also employed to further increase the performance. One of the main concerns of employing GPUs for generation of random numbers is delay, which compared to the similar computational methods including ASIC, FPGA, and physical methods such as optics may be considered as the major drawback of these relatively general purpose computational platforms. The proposed method can prove tremendous advantageous when employed on applications where slight delay is not a matter of great concern and the performance and the cost efficiency of the solution are considered. 
Our proposed methodology achieves the outstanding throughput of 2.90 Tb/s on Nvidia V 100, outperforming the Nvidia's proprietary cuRAND library while striking a notable balance in criteria of performance per cost.

\bibliographystyle{IEEEtran}
\bibliography{bare_jrnl}

\end{document}